# Forbush decrease observed by SEVAN particle detector network on November 4, 2021

A.Chilingarian [ORCID iD: 0000-0002-2018-9715], Artem Alikhanyan National Lab (Yerevan Physics Institute), Armenia, chili@aragats.am.
G. Hovsepyan, [ID:0000-0003-1027-709X', Artem Alikhanyan National Lab (Yerevan Physics Institute), Armenia.
H. Martoyan, [ORCID iD: 0000-0002-1230-8531], Artem Alikhanyan National Lab (Yerevan Physics Institute), Armenia.
T. Karapetyan, [ORCID iD: 0000-0002-0319-8433], Artem Alikhanyan National Lab (Yerevan Physics Institute), Armenia, ktigran79@gmail.com.
B. Sargsyan, Artem Alikhanyan National Lab (Yerevan Physics Institute), Armenia.
R. Langer, [ORCID iD: 0000-0002-9252-7293], Institute of Experimental Physics SAS, Košice, Slovakia.
N. Nikolova [ORCID iD: 0000-0003-1579-7801], Basic Environmental Observatory Moussala Lab (Institute for Nuclear Research and Nuclear Energy, Bulgarian Acadamy of Sciences), Bulgaria, nikol@inrne.bas.bg
Hristo Angelov, [ORCID iD: 0000-0002-2562-8924], Basic Environmental Observatory Moussala Lab (Institute for Nuclear Research and Nuclear Energy, Bulgarian Acadamy of Sciences), Bulgaria, hangelov@inrne.bas.bg
Diana Haas, Deutsches Elektronen-Synchrotron DESY, Hamburg, Germany
Johannes Knapp, Deutsches Elektronen-Synchrotron DESY, Zeuthen, Germany
Michael Walter, [ORCID iD: 0000-0001-6380-5234], Deutsches Elektronen-Synchrotron DESY, Zeuthen, Germany
Ondrej Ploc, [ORCID iD: 0000-0002-2262-4602], Nuclear Physics Institute of the CAS, v.v.i., Rez, Czechia
Jakub Šlegl, [ORCID iD: 0000-0002-4563-618X], Nuclear Physics Institute of the CAS, v.v.i., Rez, Czechia
Martin Kákona, [ORCID iD: 0000-0002-9156-0550], Nuclear Physics Institute of the CAS, v.v.i., Rez, Czechia
Iva Ambrožová, [ORCID iD: 0000-0002-2329-5783], Nuclear Physics Institute of the CAS, v.v.i., Rez, Czechia

## Abstract

On November 3-4 2021, an interplanetary coronal mass injection (ICME) hits the magnetosphere, sparking a strong G3-class geomagnetic storm and auroras as far south as California and New Mexico. All detectors of the SEVAN network registered a Forbush decrease

(FD) of 5-10% deep in 1 minute time series of count rates. We present the results of a comparison of Fd registered on mountain altitudes on Aragats (Armenia), Lomnicky Stit (Slovakia), Musala (Bulgaria), and at sea level DESY (Hamburg, Germany), and in Mileshovka, Czechia. We present as well purity and barometric coefficients of different coincidences of SEVAN detector layers on Aragats. We demonstrate disturbances of the near-surface electric (NSEF) and geomagnetic fields at the arrival of the ICME on Earth.

## 1. Introduction

The Sun is a tremendously variable object, capable of modulating the fluxes of the galactic cosmic rays (GCRs) and sending intense fluxes of solar cosmic rays (SCR) in the Earth's direction. The Sun ''modulates'' the low-energy Galactic Cosmic Rays (GCRs) in several ways. Along with broad-band electromagnetic radiation, the explosive flaring processes on the Sun usually result in Coronal Mass Ejections (CMEs) and in the acceleration of copious electrons and ions. Huge magnetized plasma structures usually headed by shock waves travel into the interplanetary space with velocities up to 3000 km/s (so-called interplanetary coronal mass ejection – ICME) and disturb the interplanetary magnetic field (IMF) and magnetosphere. These disturbances can lead to major geomagnetic storms harming multi-billion assets in space and on the ground. At the same time, these disturbances introduce anisotropy in the GCR flux. Thus, the time series of intensities of high-energy particles can provide highly cost-effective information also for the forecasting of geomagnetic storms (Leerungnavarat et al., 2003).
For the basic research of solar physics, solar-terrestrial connections, and space weather, as well as for establishing services of alerting and forecasting of dangerous consequences of space storms the networks of particle detectors located at different geographical coordinates and measuring various species of secondary cosmic rays are of vital importance.
In 1957, in an unprecedented international cooperation, more than 66.000 scientists and engineers from 67 nations perform measurements of the major geophysical parameters in the framework of the International Geophysical Year (IGY1957, Chapman, 1959).
Fifty years on, the International Heliophysical Year (IHY 2007, Thompson et al., 2009) again drew scientists and engineers from around the globe in a coordinated observation campaign of the heliosphere and its effects on planet Earth. The United Nations Office for Outer Space Affairs, through the United Nations Basic Space Science Initiative (UNBSSI), assisted scientists and engineers from all over the world in participating in the IHY. One of the the most successful projects of IHY 2007 was deploying arrays of small, inexpensive instruments around the world to get global measurements of ionospheric and heliospheric phenomena. The small instrument program was (and still is) a partnership between instrument developers and instrument hosts in developing countries. The lead scientist prepared and installed the instruments and helped to run them; the host countries provided manpower for instrument operation and maintenance. The lead scientist's institution developed joint databases, prepared tools for user-friendly access to the data, and assisted in staff training and paper writing to promote space science activities in developing countries.
A network of particle detectors located at middle to low latitudes, known as SEVAN (Space Environment Viewing and Analysis Network, Fig. 1, Chilingarian et al., 2009, 2018) was developed in the framework of the International Heliophysical Year (IHY-2007) and now operates and continues to expand within International Space Weather Initiative (ISWI). SEVAN detectors measure time series of charged and neutral secondary particles born in cascades originating in the atmosphere by nuclear interactions of protons and nuclei accelerated in the Galaxy and nearby the Sun. The SEVAN network is compatible with the currently operating high-latitude neutron monitor networks ''Spaceship Earth" (Kuwabara et al., 2006), coordinated by the Bartol Research Center, the Solar Neutron Telescopes (SNT) network coordinated by Nagoya University (Tsuchiya et al., 2013), the Global Muon Detector Network (GMDN) (Munakata et al., 2000, Rockenbach et al., 2011), and the Neutron Monitor Data Base (NMDB, Mavromichalaki et al., 2011, http://www. nmdb.eu/). The analogical detector is in operation in China (Tibet: 30.11N, 90.53E, altitude 4300 m, Zhang et al., 2010).

Three SEVAN detectors are operating in Armenia (on the slopes of Aragats Mt.: 40.25N, 44.15E, altitude 2000, 3200 m), in Croatia (Zagreb observatory: 45.82N, 15.97E, altitude 120 m), Bulgaria (Mt. Musala: 42.1N, 23.35E, altitude 2930 m), in Slovakia (Mt. Lomnicky Stit: 49.2N, 20.22E, altitude 2634 m, on Milesovka hill (50.6N, 13.9E, altitude 837 m) in Czech republic, at DESY Zeuthen (52.3N, 13.5E, altitude 30 m) and DESY Hamburg (53.5730N, 9.8810E, altitude 20m ). The potential recipients of SEVAN detectors are Italy, Israel, France, Jordan, and Algeria.

Just in the first years of the SEVAN network operation, we recognize that not only solar activity modulates the fluxes of secondary rays, but also various effects in the terrestrial atmosphere and ionosphere. During thunderstorms, emerging strong electric fields modulate the secondary particle energy spectra initiating short and extended bursts. The impulsive enhancements of the particle fluxes (so-called thunderstorm ground enhancements – TGEs, Chilingarian, et al., 2010) are disclosed as peaks in the time series of count rates of particle detectors coinciding with the strong atmospheric electric field, which accelerates and multiplies free electrons of cosmic rays. The physics of particle burst phenomena, usually connected both to the EAS phenomenon and complicated atmospheric processes, is now referred to as high-energy physics in the atmosphere (HEPA). Solar, astroparticle, and atmospheric physics are synergistically connected and need to exchange results for the explanation of particle bursts and for revealing the influence of solar flares, violent explosions in the galaxy and beyond, as well as the influence of the atmospheric electric fields on the fluxes of secondary cosmic rays registered on the Earth's surface.
The synergy of all high-energy astrophysics branches will open new areas of research for better understanding and developing physics of the geo-space. Geophysical research is becoming more and more important in the coming decades of rapidly rising natural disasters.

## 2. SEVAN detector

The basic detector of the SEVAN network (Fig. 1a, see Chilingarian et al., 2018) consists of standard slabs of 50 x 50 x 5cm$^3$ plastic scintillators. Between two identical assemblies of 100 x 100 x 5 cm$^3$ scintillators (four standard slabs) are located two 100 x 100 x 5 cm$^3$ lead absorbers, and a thick 50 x 50 x 25 cm$^3$ scintillator stack (5 standard slabs). Scintillator lights capture cones, and PMTs are located on the top, bottom and in intermediate layers of the detector. The total weight of the SEVAN detector including steel frame and detector housings is ≈1,5 tons.

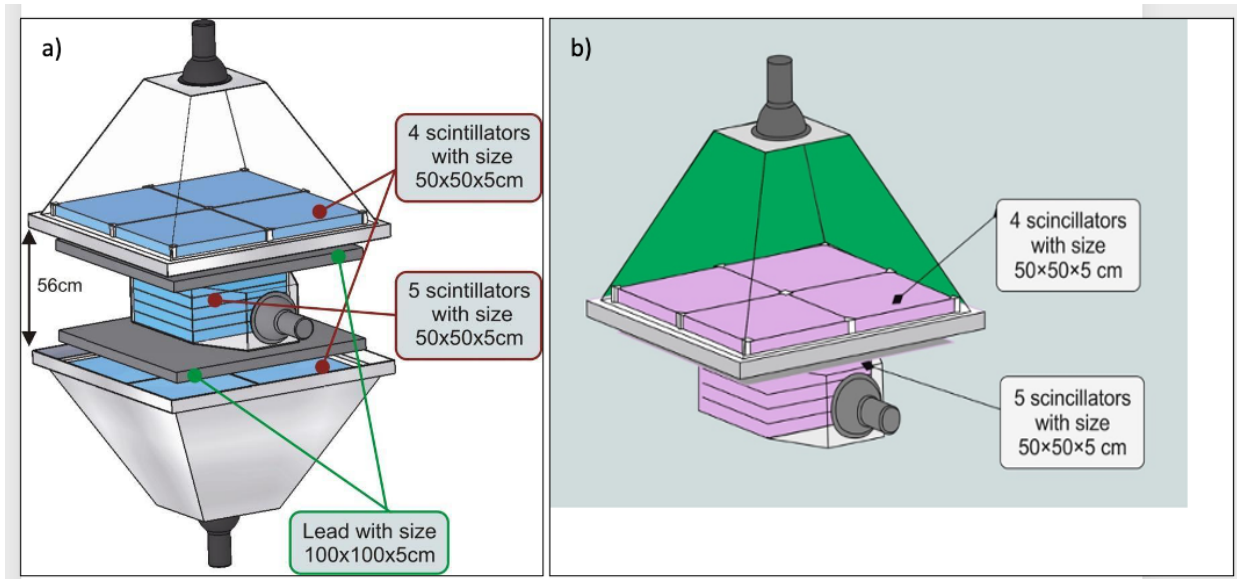

**Figure 1. Basic SEVAN detector(1a), and SEVAN-light detector (1b)**

Data Acquisition (DAQ) electronics provide registration and storage of all logical combinations of the detector signals for further offline analysis and for online alerts issuing, thus, allowing the registration of 3 species of incident particles. If we denote by ''1'' the signal from a scintillator and by ''0'' the absence of a signal, then the following combinations of the 3-layered detector output are possible: 111 and 101—traversal of high energy muon; 010—traversal of a neutral particle;100—traversal of low energy charged particle stopped in the scintillator or in the first lead absorber (energy less than ≈100 MeV). 110—traversal of a higher energy charged particle stopped in the second lead absorber. 001—registration of inclined charged particles. The Data Acquisition electronics (DAQ) allows the remote control of the PMT high voltage and of other parameters of the detector.

In 2023 we plan to install modernized SEVAN-light detectors at the Umwelt-Forschungs-Station (UFS, Schneefernerhaus, 2650 m asl, see Fig.1b) near the top of the Zugspitze (2962 m), a site with a long history of atmospheric research, where Joachim Kuettner made his seminal experiments on the structure of the electric field in the lower atmosphere (Kuettner, 1950). Due to the building constrains at UFS, SEVAN-light should be compact, shorter, and much lighter than the basic SEVAN. Thus, SEVAN-light consists only of 2 layers and the lead absorber also is not included (total weight ≈150 kg). However, we add a modernized electronics board with a logarithmic amplitude-to-digit-converter (LADC), which will provide the particle energy spectrum recovery in the range of 0.3-100 MeV. The SEVAN-light will be fully operational for high-energy atmospheric physics research, with the additional feature of measuring the energy spectrum of TGE particles. The cosmic ray variation studies, related to research in solar physics, and space weather domains will be also continued with low-energy charged and neutral particles, and their energy spectra.

In Table 1 we present the purity of the particle detector flux observed by different coincidences of the SEVAN basic detector. As we can see from the Table, the "010" coincidence efficiently selects neutrons (52%) and gamma rays (28%), "111" coincidence – muons (96%), and "100" – low energy muons and electrons (80%).

**Table 1. Purity of the SEVAN coincidences measuring ambient population of secondary cosmic ray flux (background) flux on Aragats (3200 m) in percent**

| Coinc. | neutron | proton | mu+ | mu- | $e^-$ | $e^+$ | $\gamma$ |
|---|---|---|---|---|---|---|---|
| 001 | 12.75 | 1.29 | 43.52 | 37.65 | 1.13 | 1.20 | 2.50 |
| 111 | 0.78 | 1.45 | 51.47 | 44.28 | 0.42 | 1.50 | 0.12 |
| 110 | 8.70 | 5.40 | 34.30 | 30.55 | 7.46 | 10.24 | 3.36 |
| 100 | 6.74 | 3.76 | 27.74 | 24.36 | 11.69 | 9.68 | 16.03 |
| 011 | 9.27 | 0.59 | 47.38 | 41.00 | 0.09 | 0.31 | 1.36 |
| 101 | 0.53 | 1.51 | 52.18 | 45.56 | 0.07 | 0.12 | 0.04 |
| 010 | 52.05 | 0.33 | 8.41 | 7.86 | 1.60 | 1.65 | 28.09 |

### 3. Particle flux correction to atmospheric pressure

To disentangle the atmospheric pressure and solar modulation (sometimes rather weak) effects on particle flux intensity the correction to atmospheric pressure (barometric effect) is usually made.
Barometric coefficients were calculated for the SEVAN detector on Aragats, using time series of 1-minute pressure data measured by the wireless Vantage Pro2TM plus weather station.
October 26-28 2018, data, when was 19 mb total change in the atmospheric pressure were used. The estimate of the barometric coefficient was found by the linear correlation between the intensity of the cosmic ray flux and corresponding data on atmospheric pressure (Dorman, 1975, Chilingarian and Karapetyan, 2011).
The least square method was used to obtain the regression coefficients.
In Table 2 we show the barometric coefficients for SEVAN coincidences. In the columns accordingly are posted the altitude, cutoff rigidity; barometric coefficient, goodness of fit in the form of the correlation coefficient; count rate, "Poisson" estimate of relative error (standard deviation divided by average count rate), and actual relative error. The values posted in the last two columns should be very close to each other, and as it can be seen from Table 2, the values are basically identical. This means that adopted model (linear correlation between atmospheric pressure and count rate plus Gaussian random noise) is correct.

**Table 2. Barometric coefficients, count rates, and relative errors calculated for the SEVAN detector (October 26-28 data, 2018, 19 mb total change in the atmospheric pressure)**

| Detector | Alt. (m.) | Rc (GV) | Barometric Coeff. %/mb (Oct-2018) | Correlation Coefficient | 1 min. count rate [mean] | Relative Error | $\frac{1}{\sqrt{N}}$ |
|---|---|---|---|---|---|---|---|
| SEVAN Aragats Upper 5cm | 3200 | 7.1 | -0.33 ± 0.02 | -0.986 | 29333 | 0.006 | 0.006 |
| SEVAN Aragats Middle 20cm | 3200 | 7.1 | -0.32 ± 0.02 | -0.984 | 7848 | 0.011 | 0.011 |
| SEVAN Aragats Lower 5cm | 3200 | 7.1 | -0.25 ± 0.01 | -0.981 | 17652 | 0.008 | 0.008 |
| SEVAN Aragats Coincidence 100 (low energy charged particles) | 3200 | 7.1 | -0.38 ± 0.02 | -0.984 | 20246 | 0.007 | 0.007 |
| SEVAN Aragats Coincidence 010 (mostly neutrons and gamma rays) | 3200 | 7.1 | -0.47 ± 0.04 | -0.966 | 2297 | 0.020 | 0.020 |
| SEVAN Aragats Coincidence 111 (high energy muons) | 3200 | 7.1 | -0.19 ± 0.001 | -0.966 | 3465 | 0.020 | 0.020 |
| SEVAN Aragats Coincidence 101&111 (high energy muons) | 3200 | 7.1 | -0.19 ± 0.001 | -0.966 | 7754 | 0.010 | 0.010 |
| SEVAN Aragats Coincidence 101 | 3200 | 7.1 | -0.19 ± 0.001 | -0.949 | 4289 | 0.015 | 0.015 |
| SEVAN Aragats Coincidence 110 | 3200 | 7.1 | -0.41 ± 0.03 | -0.963 | 1333 | 0.030 | 0.030 |
| SEVAN Aragats Coincidence 011 | 3200 | 7.1 | -0.27 ± 0.01 | -0.929 | 753 | 0.040 | 0.040 |
| SEVAN Aragats Coincidence 001 | 3200 | 7.1 | -0.30 ± 0.02 | -0.977 | 9144 | 0.010 | 0.010 |

In Fig.2 we show the pressure corrected and uncorrected time series of the SEVAN "010" coincidence along with the atmospheric pressure time series.

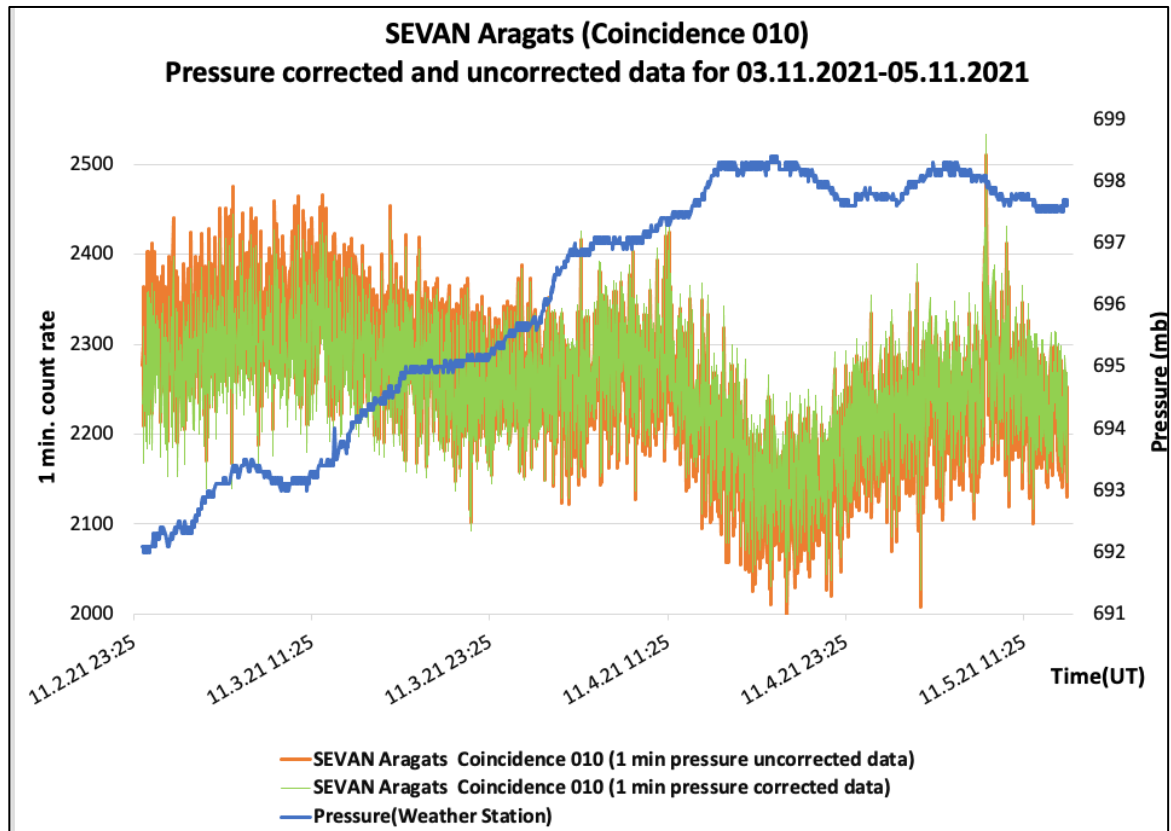

**Figure 2. One-minute time series of pressure corrected and uncorrected count rates of SEVAN "010" (neutral particles) counts. By blue curve the atmospheric pressure time series are shown.**

## 4. Forbush decrease measured by SEVAN network

The flux of low-energy Galactic cosmic rays (GCRs) is routinely modulated by the solar wind that changes the structures and polarities of the local magnetic fields in the heliosphere and magnetosphere. Thus, variations in the intensity of secondary cosmic rays observed at the Earth's surface can provide valuable information on the distribution of these structures in the heliosphere and on interactions of the solar wind with the magnetosphere.

The flux of GCRs shows short-term variations produced by the passage of the "fast" solar wind (well associated with corotating high-speed solar wind streams), shock waves, and magnetized solar plasma ejected from the Sun (i.e., the interplanetary manifestation of an interplanetary coronal mass ejections – ICME). The particle fluxes measured on the Earth's surface exhibit depletions (called Forbush decreases -FDs) and enhancement (Geomagnetic effects – GM) due to disorders in near-earth magnetic structures as a reaction to propagated shocks and ICMEs. Forbush decreases are the most frequent and easy-to-detect phenomenon of solar modulation of galactic cosmic rays. Historically, more than eighty years ago Scott Forbush was the first who related these depletions of cosmic radiation (CR) flux with solar eruptions (Forbush, 1954).

When observed with networks of particle detectors, FDs exhibit a highly asymmetrical structure: a fast decrease of the flux with a time scale of some hours, followed by a smooth recovery with a time scale of several days. If episodes of solar burst follow each over (usually from one and the same active region), several fast waves of magnetized solar plasma are travelling in the interplanetary space simultaneously, and sometimes overtake each other, thus, FDs can have rather complex structures demonstrating consequent depletions without recovery stage. Cosmic ray flux typically demonstrates pre-increases by about $1 - 2$ %, which occurs because cosmic rays are reflected on the approaching solar wind shock. When the shock wave from fast solar wind reaches Earth's magnetosphere, in most cases, an abrupt change in the geomagnetic field, named sudden storm commencement (SSC) is detected, then the main decrease phase caused by the solar winds/ejecta's southward magnetic field, and then a recovery phase finalizes a geomagnetic storm.

At the arrival of the magnetized plasma cloud to the magnetosphere, a major geomagnetic storm (GS) occurred. A measure of GS is the disturbance storm time index (DST), obtained by superposing the geomagnetic field at 3 geomagnetic locations 120 degrees longitude apart, in this way integrating the diurnal effects. During storms, DST drops below the zero level, as much as ≈500 nT in extreme cases. The magnetic field of ICME usually is resolved in 3 components, $B_x$ along the Sun-Earth line, $B_y$ perpendicular to Bx in the Earth's orbital plane around the Sun, and the $B_z$ component perpendicular to both. The $B_z$ component, parallel to the Earth's rotation axes is usually the most important in unleashing GS.

Measurements of the FD magnitude in the fluxes of different secondary CR species reveal important correlations with the speed, size of the ICME, and the ''frozen" in ICME magnetic field strength (Chilingarian and Bostanjyan, 2010). Measurements of all the secondary cosmic-ray fluxes at one and the same location are preferable due to the effects of the longitudinal dependence of the FD magnitudes (Haurwitz et al., 1965). The research of the diurnal variations of GCR by the observed fluxes of charged and neutral secondary CR also opens possibilities to correlate the changes of parameters of the daily wave (amplitude, phase, maximal limiting rigidity) with the energy of GCRs (Mailyan and Chilingarian, 2010). An example of practical application of surface CR measurements including FD is given in (Kakona et al., 2016)

On 3-5 November 2021, occurred the largest of the current 25th solar activity cycle FD; corresponding GMS unleashed auroras as far low latitude as New Mexico (39N)! SOHO coronagraphs caught the storm cloud leaving the Sun on Nov. 2, following and overtaking a slow-motion solar flare (M1.7) in the magnetic canopy of sunspot AR2891. All coincidences of the SEVAN network registered this FD by its detectors located in Aragats, Lomnicky Stit, Musala, Mileshovka, Berlin, and Hamburg. In Fig. 3 we show FD in 1-minute time series of count rates of the "010" coincidence (mostly neutrons). The FDs at mountain altitudes (Aragats, Musala, Lomnicky Stit, all above 2500 m) are pronounced rather well, better than at lower altitude (Mileshovka, ≈800 m) and at see level (Hamburg and Berlin, see Fig. 4).

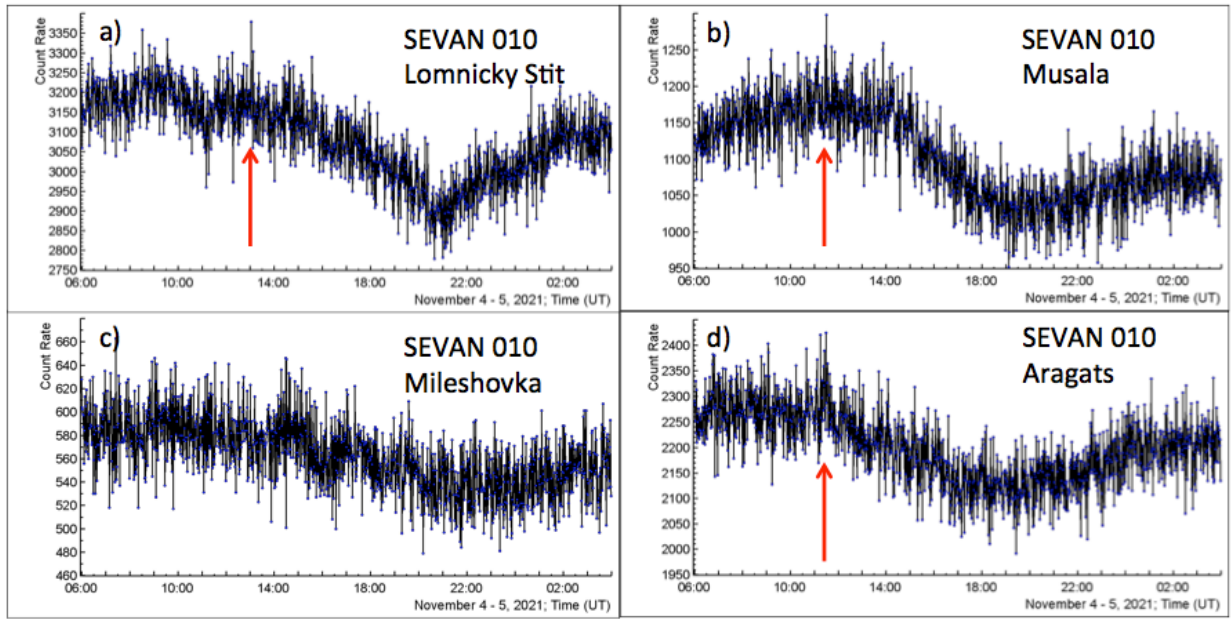

**Figure 3.** 1-minute time series of count rates of "010" coincidence of SEVAN layers (mostly neutrons), by arrows we show the "pre-Forbush" count rate enhancement followed by FD at high-altitude detectors

In Fig. 4 we show the time series of count rates of sea-level SEVAN detectors in comparison with Lomnicky Stit unite. Time series are well correlated, however, the FD pattern for sea-level monitors is smeared and its amplitude is smaller than for mountain ones.

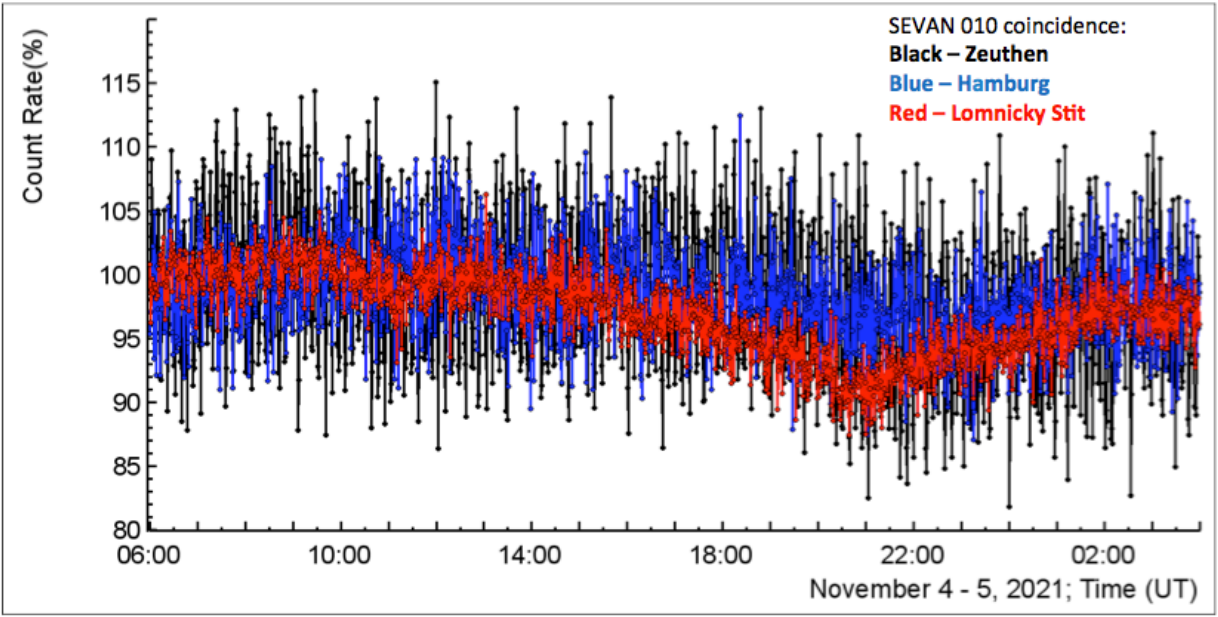

**Figure 4.** 1-minute time series of count rates of "010" (mostly neutrons) coincidence of Lomnicky Stit (red) SEVAN detectors located at sea level in Zeuthen, near Berlin (black), and in Hamburg (blue).

In Fig. 5 we compare FD registration by the Neutron Monitor and "010" coincidence of SEVAN unit, both located in the one and same place at Lomnicky Stit. The correlation of both is perfect (correlation coefficient R= 0.92), although the FD minimum measured by NM is deeper than that measured by SEVAN.

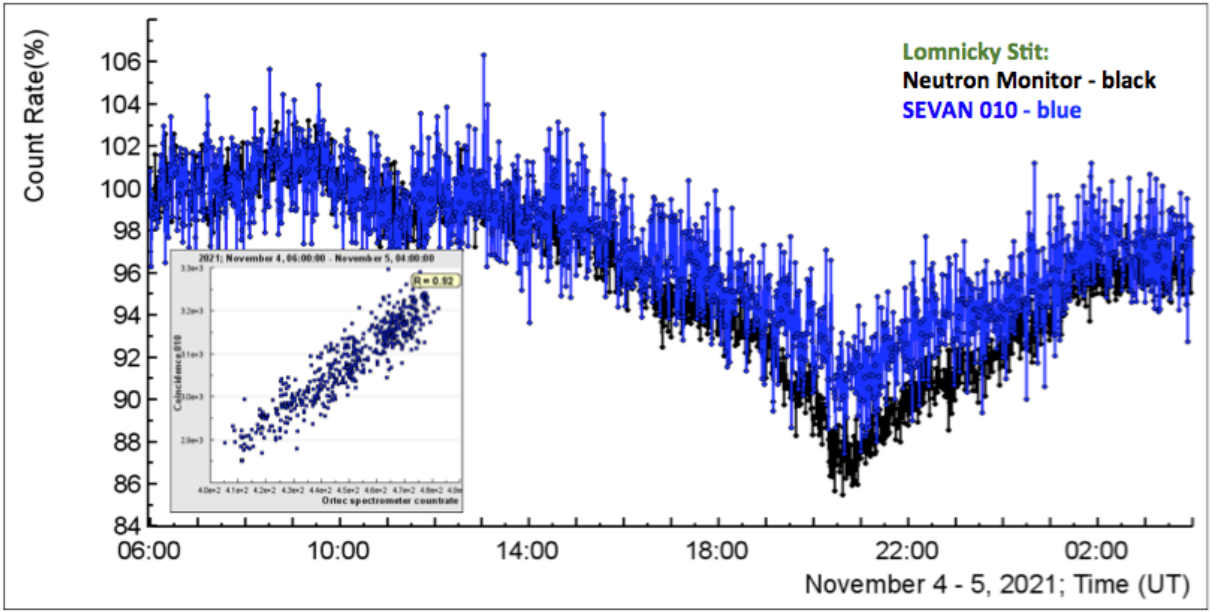

**Figure 5.** Figure 5. 1-minute time series of count rates of Neutron Monitor (black) and SEVAN's "010" (mostly neutrons) coincidence, both located at Lomnicky Stit; in the inset, we show the scatter plot.

In Fig. 6 we show the 1-minute time series of "100" coincidences of Aragats and Lomnicky Stit SEVANs. The "100" coincidence selects low-energy particles, mostly, muons, electrons, and gamma rays, see Table 1. The count rate is much larger than for the "010" coincidence and FD is pronounced much better. By green lines, we show that on Aragats minimum of FD was reached ≈1 hour before Lomnicky Stit. By red lines, we show "pre-Forbush" variations of count rates due to the interaction of galactic cosmic rays with the disturbed magnetosphere. These precursors allowing forecasting of the upcoming GMS are coinciding for both monitors.

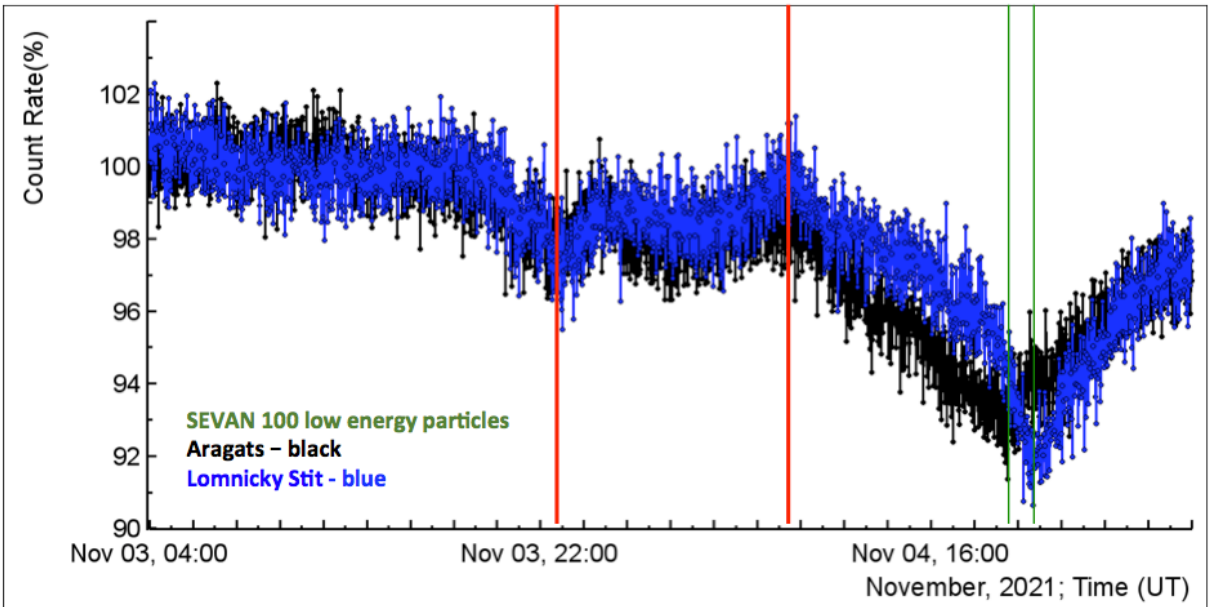

**Figure 6. 1-minute time series of count rates of "100" (low energy particles) coincidence of Aragats (black) and Lomnicky Stit (blue) SEVAN detectors. By green lines, we show the delay of FD minimum at Lomnicky Stit by ≈1 hour compared with Aragats, and by red lines – the "pre-FD" variations of cosmic ray intensity.**

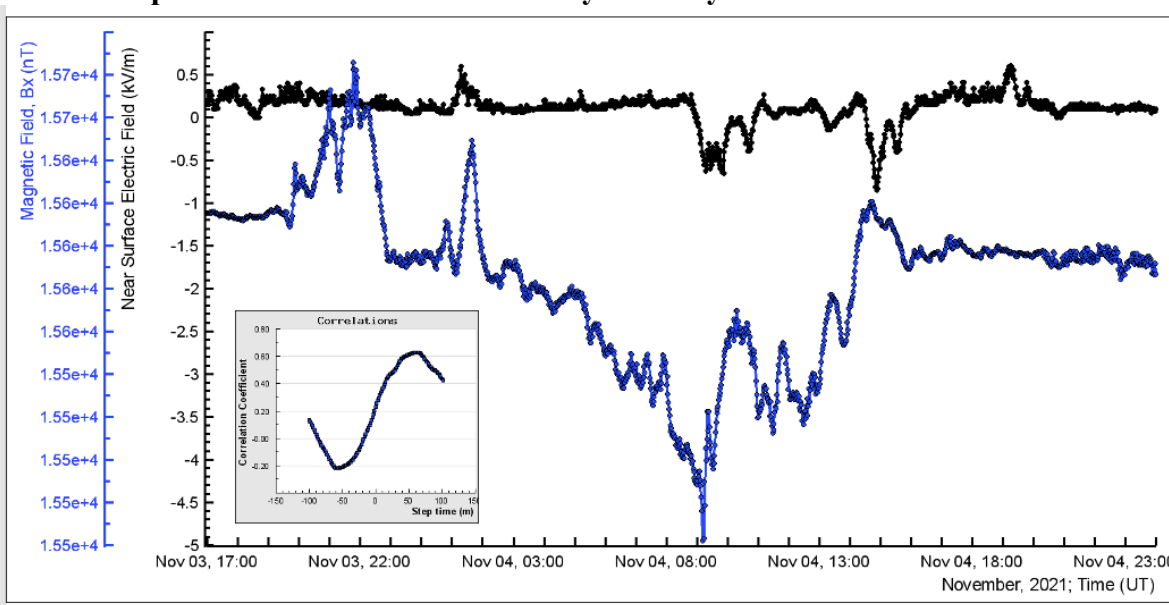

**Figure 7. Disturbances of the X component of the geomagnetic field (blue) and near-surface electric field (NSEF, black) during FD. In the inset we show the delayed**

**correlations histogram: NSEF disturbances are late relative to geomagnetic field disturbances by ≈50 minutes.**

In Fig. 7 we show measurements of the disturbances of electric and geomagnetic fields observed at the time when the combined CME reaches our planet. We show the X component of the geomagnetic field reaching a minimum value around 8:00 UT, the disturbances of the electric field are well correlated (correlation coefficient is above 0.6) with disturbances of the geomagnetic field with a delay of ≈ 50 minutes.

## 5. Conclusion

The SEVAN network measured FD in the fluxes of various species of secondary cosmic rays. The time of reaching the FD minimum was approximately the same for the mountain peak locations, and slightly different for the sea level location. The amplitude of the FD measured by SEVAN units on Aragats, Musala, and Lomnicky Stit mountains was also approximately the same, which is common for the highly isotropic FDs. At Aragats at the arrival of the magnetized plasma cloud, disturbances of geomagnetic and electric fields were observed.

The big advantage of the SEVAN network is that FD is measured in the fluxes of different particles with various energy thresholds. As we show in (see Tables Figs 5-14, and Tables 2-5 of Chilingarian et al., 2005) the correlations between FDs measured in the fluxes of different particles are well correlated with the geo-effectiveness of the solar event. For the largest in 23-rd solar activity Halloween FD (October 29, 2003) the correlation between FD parameters measured on Aragats and DST (-360 nT) reaches ≈ 1 (see Fig. 15 and Table 6 of Chilingarian et al., 2005). The collaborative efforts of SEVAN network hosts in the measurements of the solar modulation of the GCR on mountain tops in Armenia, Slovakia, Chechia, Croatia, Bulgaria, and now also in Germany were granted by new important discoveries. The 24/7 monitoring of particle fluxes with synchronized networks of identical sensors is supported by the ADEI data analysis platform (Chilingaryan, et al., 2008) that stores the multivariate data in databases with open, fast, and reliable access. The visualization and online correlation analysis of the big data coming from the SEVAN network highly improved the nowcasting and forecasting of violent solar events.

Calculated purities of the secondary cosmic ray registration for the SEVAN coincidences demonstrate the SEVAN detector's capacity to measure charged and neutral secondary cosmic ray fluxes separately, which is extremely important both for solar and atmospheric physics research. Modernization of SEVAN electronics, allowing measurement of energy spectra of neutral and charged fluxes on a minute time scale, will highly improve network abilities in research of solar modulation and atmospheric effects.